\documentclass[aps,prl,reprint,nobibnotes]{revtex4-2}

\usepackage{graphicx}
\usepackage{dcolumn}
\usepackage{bm}
\usepackage{lineno}
\usepackage{ulem}
\usepackage{soul}
\usepackage{color}
\usepackage{array}
\usepackage{booktabs}
\usepackage{multirow, makecell}
\usepackage{physics}
\usepackage[colorlinks=true, allcolors=blue]{hyperref}%

\newcommand{\Ce}{Ce$_{2}$IrIn$_{8}$}

\newcommand{\Tk}{$T_{K}$}
\newcommand{\Tc}{$T^{*}$}
\newcommand{\Ef}{$E_{F}$}

\begin{document}

\title{Orbital anisotropy of heavy fermion \Ce\ under crystalline electric field and its energy scale}

\author{Bo Gyu Jang$^{1,3}$}\thanks{These authors contributed equally to this work.}
\author{Beomjoon Goh$^1$}\thanks{These authors contributed equally to this work.}
\author{Junwon Kim$^2$}\thanks{These authors contributed equally to this work.}
\author{Jae Nyeong Kim$^1$}\thanks{These authors contributed equally to this work.}
\author{Hanhim Kang$^1$}
\author{Kristjan Haule$^5$}
\author{Gabriel Kotliar$^5$}
\author{Hongchul Choi$^4$}
\email{chhchl@gmail.com}
\author{Ji Hoon Shim$^{1,2}$}
\email{jhshim@postech.ac.kr}

\affiliation{$^1$Department of Chemistry, Pohang University of Science and Technology, Pohang 37673, Korea} 
\affiliation{$^2$Division of Advanced Materials Science, Pohang University of Science and Technology, Pohang 37673, Korea}
\affiliation{$^3$Korea Institute for Advanced Study, Seoul 02455, Korea}
\affiliation{$^4$Center for Correlated Electron Systems, Institute for Basic Science (IBS), Seoul 08826, Korea}
\affiliation{$^5$Department of Physics and Astronomy, Rutgers University, New Jersey 08854, USA}

\begin{abstract} 

We investigate the temperature ($T$)-evolution of orbital anisotropy and its effect on spectral function and optical conductivity in \Ce\, using a first principles dynamical mean field theory combined with density functional theory. The orbital anisotropy develops by lowering $T$ and it is intensified below 
a temperature corresponding to the crystalline-electric field (CEF) splitting size.
Interestingly, the depopulation of CEF excited states leaves a spectroscopic signature, ``shoulder", in the $T$-dependent spectral function at the Fermi level.
From the two-orbital Anderson impurity model, we demonstrate that CEF splitting size is the key ingredient influencing the emergence and the position of the ``shoulder". Besides the two conventional temperature scales \Tk\ and \Tc , we introduce an additional temperature scale to deal with the orbital anisotropy in heavy fermion systems.

\end{abstract}


\maketitle

The two characteristic temperatures, the Kondo temperature, \Tk\ and the coherence temperature, \Tc\ have successfully been used to describe the electronic structures of Ce-based heavy fermion compounds \cite{Shim2007, Haule2010, Choi2012, Kang2019}. The local Kondo temperature \Tk\ is identified from the $T$-evolution of the area of a Fermi surface (FS), which starts to change at the onset of the hybridization \cite{Burdin2000} while the coherence temperature \Tc\ is defined from the lattice coherence of quasiparticle states, showing the coincidence with the resistivity maximum \cite{Choi2012, Kang2019, Jang2020}. The quasiparticle relaxation rates measured in ultrafast optical spectroscopy signalled two characteristic temperatures corresponding to \Tk\ and \Tc\ \cite{Liu2020}. Angle-resolved photoemission spectroscopy (ARPES) measurements of Ce-based heavy fermion compounds showed that the evolution of FS was almost completed at \Tc\ \cite{Yao2019, Klotz2018, Liu2019, Chen2017, Chen2018}. The quasiparticle weights would keep enhanced within the framed FS below \Tc. Both temperature scales seem quite successful to capture the key events in heavy fermion compounds which happen amid the temperature change. But it's not clear whether both reflect the influences of the crystalline-electric field (CEF). 

In recent years, there has been an increasing interest in the role of orbital anisotropy in the heavy fermion systems, thanks to the development of non-resonant inelastic X-ray scattering (NIXS) technique and high-resolution in ARPES. T. Willer \textit{et al}. claimed that the ground state of Ce$M$In$_{5}$ ($M$ = Co, Rh, and Ir) depends on the orbital anisotropy of Ce ion driven by each CEF environment \cite{Willers2015}. They found that the superconducting materials, CeCoIn$_{5}$ and CeIrIn$_{5}$, have prolate shaped 4$f$ CEF ground state orbitals, while magnetically ordered CeRhIn$_{5}$ has an oblate shaped ground state orbital. The Ir substitution for Rh in CeRhIn$_{5}$ clearly shows that the ground state is strongly associated with the orbital shape revealing the existence and the importance of anisotropic hybridization of Ce 4$f$ electrons in the Kondo lattice materials \cite{Willers2015}. On the other hand, the recent ARPES measurements on Ce$_{m}M_{n}$In$_{3m+2n}$ indicated the continuous change of spectral weight at the Fermi level (\Ef) 
from much higher temperature than \Tc\ \cite{Jang2020, Rodolakis2018, Chen2018PRL}. In this systems, the thermal broadening effect would surpass CEF excitation energy. Therefore, incoherent CEF states would play an important role and modify the orbital anisotropy. So far there is no elaboration on any $T$-scales associated with CEF and $T$-evolution of the orbital anisotropy.

CeCu$_{2}$Si$_{2}$ also receives great attention due to the double dome shape superconducting region and its possible relation with the orbital transition originated from different CEF ground state \cite{Pourovskii2014}. 
However, later NIXS measurement demonstrated the absence of the orbital transition under pressure \cite{Rueff2015}. Since the CEF excited state is located at well above the ground state ($>$ 30 meV $\sim$ 350 K) in CeCu$_{2}$Si$_{2}$ \cite{Ehm2007, Amorese2020}, it would be hard to detect the impact of the CEF excited states on the low energy state \cite{Amorese2020}. Nonetheless, the possible multiorbital singlet paring in superconductivity was suggested \cite{Nica2021}.

Here, we have chosen \Ce\ as a test-bed material for the investigation on the temperature-driven development of the orbital anisotropy. First, we have confirmed whether our DFT+DMFT is successful to describe the experimentally known CEF states. The excited CEF states can be signalled from optical conductivity calculations.
After estimating the two conventional characteristic temperature scales, \Tk, and \Tc\, we have focused on an additional $T$-scale associated with CEF splitting. To analyze this new $T$-scale quantitatively, we have studied a two-orbital Anderson impurity model with various CEF splittings. We demonstrated that the $T$-scale where the orbital anisotropy intensifies depends on the the size of CEF splitting and it is clearly distinguished from \Tc\ and \Tk. And we were able to interpret the spectroscopic observation, ``shoulder", 
as a clue for the well-developed orbital anisotropy.

\begin{figure} 
	\centering
	\includegraphics[width=0.9\linewidth]{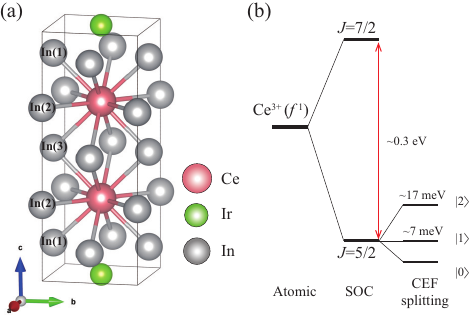}
	\caption{(a) The crystal structure of \Ce. There are two out-of-plane In(1), In(3) types and one in-plane In(2) type with respect to Ce atoms. (b) The schematic energy diagram for CEF states of Ce 4$f$ orbital based on DFT+DMFT calculation} 
	\label{fig:Fig1}
\end{figure}

Figure~1(a) displays the crystal structure of \Ce. It crystallizes with alternative stacks of one IrIn$_{2}$ and two CeIn$_{3}$ layers. There are three different types of In atoms: In(2) in Ce plane, In(1) and In(3) out of the Ce plane. The two equivalent Ce atoms are enclosed by sixteen In atoms. Four In(3) atoms are shared with the equivalent Ce atoms.  Figure 1(b) demonstrates how $J = 5/2$ states under the tetragonal symmetry are split into three doubly degenerated CEF states ($\Gamma_{6}$ , $\Gamma_{7}^{1}$, and $\Gamma_{7}^{2}$ ) \cite{Fischer1987, Settai2007}. The tetragonal CEF strengths in Ce$_{2}M_{1}$In$_{8}$ are experimentally estimated to $5\sim7$ meV and $17\sim25$ meV \cite{Ohishi2009, Yamashita2011, Malinowski2003, Willers2010}. Recent magnetic susceptibility study on \Ce\ proposed the CEF energy of 6 meV and 18 meV for the 1st and 2nd excited states, respectively \cite{Christovam2020}.

The momentum ($k$)-resolved spectral function $A(\omega, k)$ at $T=$ 10 K clearly shows the CEF splitting as shown in Fig. 2(a).  The three flat quasiparticle states centered at 0, 7, and 17 meV (indicated by the ground state $\ket{0}$, the first-excited CEF state $\ket{1}$, and the second-excited CEF states $\ket{2}$, respectively) show good agreements with the experimental results \cite{Christovam2020}. 
The black arrows in Fig. 2(a) indicate possible optical transitions, which will be discussed later.
The $k$-integrated spectral function of Ce 4$f$ states at different temperatures are illustrated in Fig. 2(b).
Although the spin-orbit coupling (SOC) splitting ($\Delta_{SO} \sim$ 0.3 eV, indicated by black arrows) between $J=5/2$ and $J=7/2$ states is clearly visible, the CEF energy splittings of the $J=5/2$ states are not clearly recognizable in the given energy window, even at $T$ = 20 K.

\begin{figure} 
	\centering
	\includegraphics[width=\linewidth]{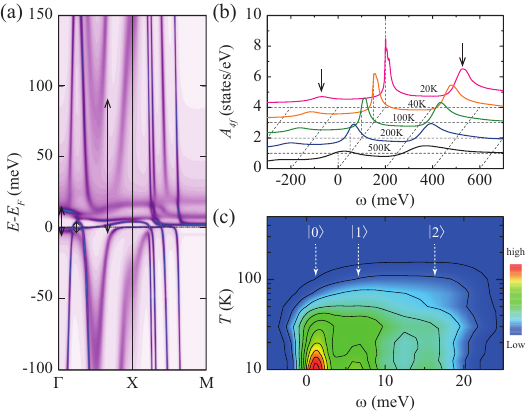}
	\caption{$k$-dependent spectral function near \Ef\ at $T$=10 K. There are flat bands from each CEF multiplet states. The arrows describe the possible optical transitions by the CEF energy splittings of 7 meV, 17 meV, and the hybridization gap of 80 meV. (b) $k$-integrated spectral function of Ce 4$f$ state for various temperatures. In addition to the central Kondo peak at \Ef\, there are two side Kondo peaks, indicated by arrows, due to the SO splitting of $J=5/2$ and $J=7/2$ states. (c) Stack plot of $T$-dependent sequential merging of the $J=5/2$ CEF states.}  
	\label{fig:Fig3}
\end{figure}

The way the CEF states are developed with lowering temperature can be identified in the very proximity of \Ef. 
The color in Fig.~2(c) indicates the quasiparticle peak intensity near the \Ef. When the temperature is higher than the CEF energy scale, for example, over 200 K ($\sim 17$ meV), all the CEF features are consolidated due to the thermal broadening effect. As the temperature decreases, the green region appears over blue background around 100 K reflecting the enhancement of the Kondo resonance. Below 50 K, two parts are developed separately.  
The higher energy part corresponds to $\ket{2}$ around 15 meV. The lower energy region undergoes another separation into $\ket{0}$ and $\ket{1}$ below 20 K. The contribution of each CEF states to the Fermi level, however, is not evident in the Fig. 2(c). This will be discussed in Fig. 3(d).

Tracking the optical transitions as a function of $T$ also allows us to visualize the sequential building-up of the CEF states, which is hardly seen in ARPES measurement. The two short black arrows in Fig.~2(a) indicate possible optical transitions between the CEF states. The hybridization gap near the $\Gamma-X$ path (See the long black arrow in Fig. 2(a)) will produce an optical transition of about 80 meV, which contains a considerable non-$f$ contribution. When the temperature is lowered, the peak at $\sim$ 80 meV starts to evolve around 200 K, signalling the onset of Ce 4$f$ contribution to FS (Figure 3(a) and SI). Therefore, the track of the peak at 80 meV would lead to an estimation of \Tk. An additional shoulder feature around 20 meV appears when the temperature is further lowered. This is a result of the optical transition between $\ket{0}$ state and $\ket{2}$ state. This shoulder feature was clearly seen in the previous experimental optical conductivity data of CeCoIn$_{5}$ at 30 meV \cite{Singley2002}, which agrees well with the energy level of $\ket{2}$.
The lowest peak in energy at 7 meV is visible below 20 K only after excluding the Drude part
from the optical conductivity as seen in the lower panel of Fig. 3(a). And the high $T$ optical conductivity has a huge Drude peak while it becomes narrower and shows a diverging behaviour below a certain temperature. 
The temperature is often considered as the coherence temperature, \Tc\ $\sim 50$ K.
The narrow Drude peak is a typical feature reported in HF systems \cite{Varma1985, Millis1987, Scheffler2005}.

As the temperature decreases, a crossover of $4f$ electrons from ``localized" to ``delocalized" occurs. Correspondingly, the Fermi surface also evolves from ``small" ones to ``large" ones. According to the Luttinger's theorem, the volume of FS can be mapped into the number of electrons which are occupied.
So the $T$-evolution of FS means the change in the number of valence electrons. Since two Ce$^{3+}$ ions exist in the unit cell, two Ce 4$f$ electrons could finally be added to the valence states. It is informative to see in Fig. 3(b) how fast the localized electrons are transferred into FS as a function of $T$.
Three regions are distinct.: the gradual increase region (high $T$), the fast linear increase region, and the saturated region (low $T$). The first crossover temperature around 200 K could be regarded as \Tk\
and it well agrees with the estimation from the optical conductivity. After the linear increase, the curve almost reaches to its maximum value around 50 K, where the 4$f$ states become quite similar to quasiparticles described by the conventional DFT calculation (itinerant 4$f$ states). This crossover indicates that most of the 4$f$ electrons become delocalized. So the two crossover temperatures, \Tk\, and \Tc\ of \Ce\ could be estimated as around 50 K and 200 K. The $T$-dependent magnetic part of resistivity $\rho_{mag}$ ($\rho_{Ce_{2}IrIn_{8}}-\rho_{La_{2}IrIn_{8}}$) makes the above estimation of \Tc\ more reliable.
The DFT+DMFT calculation (line) well reproduces the experimental resistivity (red dots) as shown in Fig.~3(c) \cite{Ohara2003}. The maximum resistivity around $\sim$45 K shows a good agreement with the second crossover temperature, \Tc\ (see the vertical guideline through Fig.~3(b-d)).

\begin{figure} 
	\centering
	\includegraphics[width=1.0\linewidth]{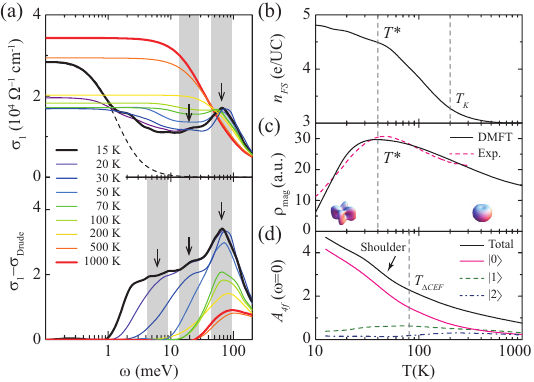}
	\caption{	
	(a) The calculated conductivities $\sigma_{1}(\omega)$ as a function of $T$. 
	The dashed lines indicates the Drude spectrum. 
	At high $T$, it shows only the broad Drude peak 
	while the inter-band transition spectrum emerges as decreasing $T$. 
	The lower figure shows the effective inter-band transition obtained by subtracting the Drude part from $\sigma_{1}(\omega)$. 
	The gray regions with arrows indicate the optical peak from the CEF energy splitting and the hybridization gap shown in Fig.~2(a). 
	(b) The calculated $T$-dependent number of valence electrons ($n_{FS}$).
	(c) The calculated (black line) and experimental (red dots) magnetic part of resistivity (4$f$ contribution). 
	(d) The $T$-evolutions of $A(\omega=0)$ for total (black line) and the CEF states. Dashed lines indicate the characteristic temperature scales (\Tc, \Tk) and $\ket{1}$ state energy scale ($T_{\Delta CEF}$).
     }  
	\label{fig:Fig4}
\end{figure}

It seems that the total spectral weight at \Ef\ via Kondo resonance just monotonically increases with decreasing the temperature, but there are interesting changes going on beneath the surface. Figure~3(d) presents the calculated spectral weight, $A_{4f}(\omega=0)$ of $\ket{0}$(red solid), $\ket{1}$(dashed green) and $\ket{2}$(dashed blue) states at \Ef. At very high $T$ ($>$ 300 K), the CEF states evenly contribute to $A(\omega=0)$ and all increase monotonically as the temperature decreases. But below $\sim$ 300 K, the contribution of $\ket{0}$, $\ket{1}$ and $\ket{2}$ states at \Ef\ become different, indicating an orbital anisotropy starts to develop. $A_{4f}(\omega=0)$ of $\ket{2}$ peaks at $\sim$250 K (blue arrows in Fig. 3(d)) and then falls. And the $A_{4f}(\omega=0)$ of $\ket{1}$ peaks secondly at $\sim$ 70 K 
(red arrows in Fig. 3(d)) and then falls, too. Note that these temperature scales are on the same order of magnitude as each CEF splitting energy, indicating that the thermal broadening surpasses the CEF splitting at higher temperature. Only the $A_{4f}(\omega=0)$ of $\ket{0}$ continues to increase.
Let us consider the two main contributions to \Ef\ given by $\ket{0}$ and $\ket{1}$ states.
After the onset of the orbital anisotropy below 300K, the orbital anisotropy first evolves slowly and it intensifies at temperatures below 70 K due to the downturn of the spectral weight of the $\ket{1}$ states.
(The two representative shapes of the $4f$ wavefunctions at \Ef\ before and long after the onset of the orbital anisotropy are shown in the inset of Fig.~3(c).) Seemingly, the temperature where the orbital anisotropy intensifies corresponds neither to the Kondo temperature \Tk\ nor the coherence temperature \Tc .

\begin{figure} 
	\centering
	\includegraphics[width=1\linewidth]{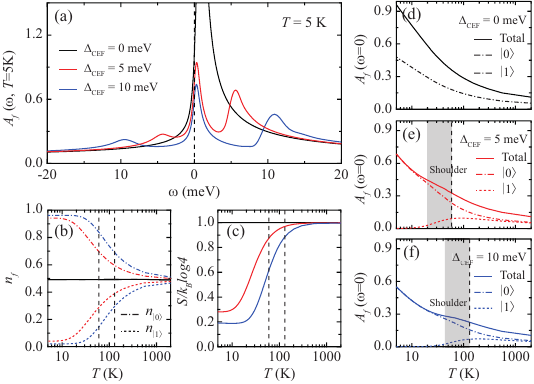}
	\caption{(a) Total impurity spectral functions $A_{f}(\omega)$ at $T$ = 5 K 
	with three different CEF energy splittings ($\Delta_{CEF}$ = 0, 5, 10 meV) 
	of two-orbital infinite-$U$ Anderson model with bare half bandwidth of 1 eV.  
	(b) $T$-evolution of occupancies for each orbitals ($n_{\ket{0, 1}}$). 
    (c) $T$-evolution of the entropy ($S$) for each CEF strength.
	(d-f) $T$-evolution of spectral function at $\omega=0$ 
	for the ground state $\ket{0}$ and the first excited state $\ket{1}$.
	Corresponding $\Delta_{CEF}$ scales, 58 K and 116 K, are drawn in dashed lines.	
	Gray regions indicate the ``shoulder" feature originated from the depopulation of $\ket{1}$ state by lowering $T$.
	}  
	\label{fig:Fig4}
\end{figure} 

One important feature that should be noted is the existence of a ``shoulder" in the $T$-evolution of the total spectral weight at E$_F$ of Fig.~3(d), which is also observed in the recent ARPES studies
on Ce$M$In$_{5}$ compounds \cite{Jang2020, Chen2018, Chen2018PRL}. Since the ``shoulder" seems to happen as a result of the depopulation of $\ket{1}$ states with lowering temperature, we have examined the role of the size of the CEF energy splittings on the ``shoulder" feature using the two-orbital infinite-$U$ Anderson impurity model. (Considering the negligible contribution, we excluded the $\ket{2}$ states.) In the square lattice, there is a conduction state of a bare half bandwidth of 1 eV. Two impurity orbitals are located at $\epsilon_{1}$= -1 eV and $\epsilon_{2}$= $\epsilon_{1}+\Delta_{CEF}$ ($\Delta_{CEF}$=0, 5, 10 meV). Hybridization strengths ($V$) is presumably a constant: $V$= 0.24 eV. The impurity model is solved using the non-crossing approximation.

Figure 4(a) shows the Kondo resonance peaks of the impurity states at $T$ = 5 K with respect to the three different $\Delta_{CEF}$ values. For $\Delta_{CEF}=0$ meV, the two degenerate impurity levels bring about 
one Kondo resonance peak only near $\omega$=0. The nonzero $\Delta_{CEF}$ drives two distinct resonance peaks above the $E_F$. And additional excitations by CEF are also produced below the $E_{F}$ as is seen 
by the spin-orbit coupling. At low temperature, the spectral weight at \Ef, $A_{f}(\omega=0)$, decreases as the CEF splitting increases. Above CEF temperature scale ($T_{\Delta CEF}$), however, $A_{f}(\omega=0)$ becomes similar for all cases, regardless of $\Delta_{CEF}$ size. As a result, $A_{f}(\omega=0)$ decreases slowly depending on temperature in the presence of nonzero CEF splitting \cite{Jang2020}.

Fig.~4(b) provides the $T$-evolution of occupancies ($n_{f}$) for each orbital. The total occupancy ($n_{f,tot}$ = $n_{f,1}$+$n_{f,2}$) remains almost constant at $\sim$0.98 for whole temperature range. They are equally populated at very high temperature. The equally occupied of $\ket{0}$ and $\ket{1}$ states start to deviate from 
the median value, 0.5, as the temperature decreases. The $T$-evolution curve is strongly influenced by $\Delta_{CEF}$. As $\Delta_{CEF}$ is larger, the deviation from the median value becomes faster. This feature is also well reflected in the $T$-evolution of the entropy ($S$). ($S$ is calculated from the occupancies of impurity orbitals, See Supplementary Material) The larger the CEF state splits, the faster the entropy drops.

It is evident from Fig.~4(d) that any ``shoulder" feature is absent when there is no $\Delta_{CEF}$. 
Since the $\ket{0}$ and $\ket{1}$ states are equally populated at all temperatures, the orbital anisotropy is also absent. However, the ``shoulders" behavior are clearly manifested when there is nonzero $\Delta_{CEF}$
as shown in Fig.~4(e,f). The orbital anisotropy is also initiated from the deviation between 
the $\ket{0}$ and $\ket{1}$ states. As the difference between $\ket{1}$ and $\ket{1}$ states gets bigger,the orbital anisotropy become mature. Seemingly, the downturn of the $\ket{1}$ state with lowering temperature
happens around the temperature corresponding to $\Delta_{CEF}$, 5 meV ($T_{\Delta CEF}$ = 58 K) and 10 meV ($T_{\Delta CEF}$ = 116 K) respectively (dashed vertical lines).The entropy at the $T_{\Delta CEF}$ in both cases was identified consistently as $\sim$ 0.9$k{_B}log4$. The ``shoulder" feature occurs below $T_{\Delta CEF}$ after substantial depopulation of the $\ket{1}$ state. It is a signature accompanied by the mature development of orbital anisotropy. Indeed, the ``shoulder" position has a proportional relationship with $T_{\Delta CEF}$.

Based on the results of the two-orbital model and the DFT+DMFT, we summarized our findings.
(1) Around $T_{\Delta CEF}$, the first excited state $\ket{1}$ contribution on $A_{f}(\omega=0)$ starts to decrease, accelerating the orbital anisotropy. This leads a ``shoulder" behavior in $T$-dependent $A_{f}(\omega=0)$ at a slightly lower temperature. 
(2) Therefore, the recently observed  ``shoulder" behaviors in ARPES experiments on Ce$M$In$_{5}$ \cite{Jang2020, Chen2018, Chen2018PRL} are the spectroscopic fingerprints of the orbital anisotropy development. The position of ``shoulder" from ARPES measurement is roughly the same order of magnitude as the first CEF excited state of Ce$M$In$_{5}$.
(3) $T_{\Delta CEF}$ and the ``shoulder" fall onto neither \Tk\ nor \Tc\ in \Ce.
This implies that the two conventional temperature \Tc\ and \Tk\ would not be available to scale the onset of orbital anisotropy.  
Therefore, it is natural to use $T_{\Delta CEF}$ as a new temperature scale to describe the orbital anisotropy in heavy fermion compounds.
(4) The order of characteristic temperature scales in \Ce\ is as follows: \Tc\ $<$ $T_{\Delta CEF}$ $<$ \Tk. Since $\Delta_{CEF}$ is smaller than \Tk, excited CEF states can affect $T$-evolution of Kondo resonance state and the orbital anisotropy develops during the ``localized" to ``delocalized" crossover. On the other hand, $T_{\Delta CEF}$ ($>$ 30 meV $\sim$ 350 K) is larger than \Tk\ in CeCu$_{2}$Si$_{2}$ \cite{Ehm2007, Amorese2020}. In this case, the excited CEF state's contribution to the Kondo resonance state is very small \cite{Amorese2020}.
(5) Finally, the optical conductivity would play an important role to reveal the $T$-evolution of a Kondo resonance state with CEF states.

\section{Acknowledgements}
We thank Prof. Jungseek Hwang and Dr. Jae Hyun Yun for useful discussion on the extended Drude analysis, Dr. J. D. Denlinger for fruitful discussion on the CEF effect. This work was supported by the National Research Foundation of Korea (NRF) grant funded by the Korea government (MSIP) (No.2015R1A2A1A15051540), the Institute for Basic Science in Korea (Grant No. IBS-R009-D1) and the Supercomputing Center/Korea Institute of Science and Technology Information with supercomputing resources including technical support (KSC-2016-C1-0003). BGJ supported by a KIAS individual Grant QP081301 at Korea Institute for Advanced Study.

\bibliography{Ce218CEF}

\end{document}